\documentclass[a4paper, conference]{IEEEtran}
\usepackage{cite}
\usepackage{amsmath,amssymb,amsfonts}
\usepackage{algorithmic}
\usepackage{graphicx}
\usepackage{textcomp}
\usepackage{xcolor}
\def\BibTeX{{\rm B\kern-.05em{\sc i\kern-.025em b}\kern-.08em
    T\kern-.1667em\lower.7ex\hbox{E}\kern-.125emX}}

\usepackage[hidelinks]{hyperref}
\usepackage[inline]{enumitem}
\usepackage{multirow}

\begin{document}
\bstctlcite{BSTcontrol} 

\title{SVFF: An Automated Framework for SR-IOV Virtual Function Management in FPGA Accelerated Virtualized Environments\\
}

\author{\IEEEauthorblockN{Stefano Cirici}
\IEEEauthorblockA{\textit{Virtual Open Systems SAS} \\
Grenoble, France \\
s.cirici@virtualopensystems.com}
\and
\IEEEauthorblockN{Michele Paolino}
\IEEEauthorblockA{\textit{Virtual Open Systems SAS} \\
Grenoble, France \\
m.paolino@virtualopensystems.com}
\and
\IEEEauthorblockN{Daniel Raho}
\IEEEauthorblockA{\textit{Virtual Open Systems SAS} \\
Grenoble, France \\
s.raho@virtualopensystems.com}
}

\maketitle

\begin{abstract}

FPGA accelerator devices have emerged as a powerful platform for implementing high-performance and scalable solutions in a wide range of industries, leveraging their reconfigurability and virtualization capabilities.
Virtualization, in particular, offers several benefits including improved security by resource isolation and sharing, and SR-IOV is the main solution for enabling it on FPGAs.

This paper introduces the SR-IOV Virtual Function Framework (SVFF), a solution that aims to simplify and enhance the management of Virtual Functions (VFs) on PCIe-attached FPGA devices in Linux and QEMU/KVM environments, solving the lack of SR-IOV re-configuration support on guests.
The framework leverages the SR-IOV support in the Xilinx Queue-based Direct Memory Access (QDMA) to automate the creation, attachment, detachment, and reconfiguration of VFs to different Virtual Machines (VMs).
A novel pause functionality for the VFIO device has been implemented in QEMU to enable the detachment of VFs from the host without detaching them from the guest, making reconfiguration of VFs transparent for guests that already have a VF attached to them without any performance loss.
The proposed solution offers the ability to automatically and seamlessly assign a set of VFs to different VMs and adjust the configuration on the fly.
Thanks to the pause functionality, it also offers the ability to attach additional VFs to new VMs without affecting devices already attached to other VMs.
\end{abstract}


\section{Introduction}

In recent years, the widespread adoption of virtualization technologies has led to a growing demand for effective and efficient management of hardware resources in data centers, cloud computing, and high-performance computing environments.
The management of devices in these environments can be challenging, especially when dealing with multiple VMs and frequent reconfigurations.
To accelerate complex workloads, Field Programmable Gate Arrays (FPGAs) are more and more used to provide a fast-deployment, fast-prototyping solution to custom problems, with an increasing interest in Artificial Intelligence (AI) and Machine Learning (ML) domains.
Virtualization of FPGA resources also enables scenarios where a single FPGA can provide multiple devices to different VM, to enhance the workload and security.

Addressing some of the VMs' guest acceleration challenges is already possible using the Single Root I/O Virtualization (SR-IOV) technology to allow multiple VMs to share the same physical device providing direct access to the hardware with minimal to no overhead.
However, SR-IOV usage is not always straightforward and can lead to complex scenarios, especially when there is the need to change the VF configuration on the fly which requires the removal of every VF from each VM.
This is not a problem in applications with fixed and predictable workloads or with dedicated resources for the VMs, but in environments such as cloud computing, software as a service, high-performance computing, and edge computing where workloads can vary rapidly, resources need to be shared among different VMs, and devices need to be reconfigured and reassigned, the limitations of SR-IOV become more evident.
In these cases, a solution that allows automatic assignment and reconfiguration of virtual functions, can significantly simplify device management and improve system efficiency.

\begin{figure}
\centering
\includegraphics[width=\columnwidth]{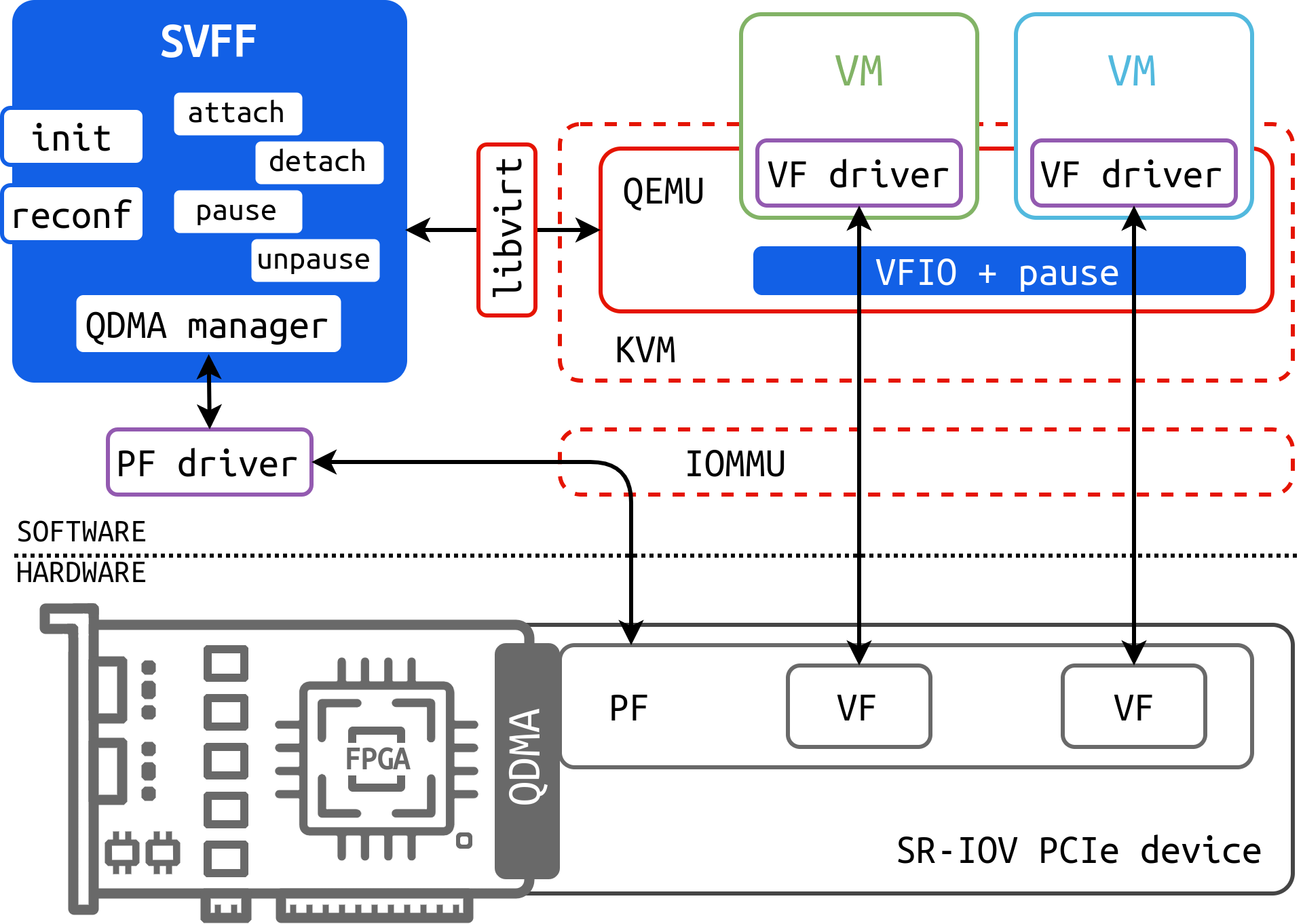}
\caption{SR-IOV Virtual Function Framework}
\label{fig:sriov}
\end{figure}

The proposed SR-IOV Virtual Function Framework (SVFF), in blue in figure \ref{fig:sriov}, leverages the support of SR-IOV in the Xilinx QDMA IP to automate the creation, attachment, detachment, and reconfiguration of accelerators to different VMs.
More in detail, SVFF is a KVM-based hardware-software framework where the hardware part is entirely managed by the Xilinx QDMA and the software improves flexibility and usability with a set of scripts that automates several operations, as well as with the implementation of a novel pause functionality that minimizes guests accelerator downtime by enabling VFs reconfiguration without detaching them from the guest.

This comes with
\begin{enumerate*}[label=(\arabic*)]
    \item native performance since SVFF is able to achieve performance comparable to standard SR-IOV accelerators (benchmarks in Section \ref{evaluation_and_results} show a 2\% reduction of delay times with SVFF);
    \item better software programmability, thanks to the pause functionality that enables transparent changes of the VF number with virtually no downtime at the guest level and with no need of modifying the guest driver;
    \item flexibility of changing the number of VF seamlessly without worrying about all the SR-IOV, drivers, and VFIO procedures needed to attach and detach them to VMs;
    \item high extensibility at the hardware level, since QDMA is a standard IP that easily enables the development of accelerators.
\end{enumerate*}

\section{Background}

This section details the key technologies beyond SVFF, starting from a brief introduction of FPGAs, and later explaining key concepts of SR-IOV, QEMU/KVM, and libvirt.

\subsection{FPGAs}

Moving away from their initial purpose as prototyping devices, FPGAs are becoming progressively more attractive for industrial and consumer solutions, also thanks to silicon technology improvements.
With the future being in parallel computing, AI, and ML, FPGAs saw the introduction in the HPC market to offer some significant performance advantages with respect to other solutions, especially when flexibility is one of the main targets \cite{moore,highperffpga}.

Scaling performance to parallelize the workload is however not as easy as just increasing the number of processors or computing units in a system.
FPGA provides the means to realize almost any custom-hardware configuration with top-class performance and low power while maintaining the flexibility of software re-programmability.
With the help of virtualization, FPGA resources can then be assigned to different and independent users (VM) to parallelize workloads in HPC scenarios.
One popular approach for exposing FPGAs to virtual machines is SR-IOV, which enables VMs to directly access physical hardware devices by assigning a portion of the hardware resources to each VM.

\subsection{SR-IOV Technology Overview} \label{sriov_technology_overview}

The SR-IOV specification enables a PCI Express (PCIe) device to appear as multiple, distinct physical devices, each with its own set of independent I/O resources \cite{sriovprimer}.
This is achieved by partitioning a Physical Function (PF) device into virtual functions (VFs), which are then allocated to each VM.
As a result, each VM can have direct access to its own set of I/O resources, without interfering with other VMs.
The Physical Function (the single-root function) is a full PCIe function that provides the primary functionality of the device.
The Virtual Function is a lightweight PCIe function (lacking some of the configuration resources) associated with the device's PF and shares one or more physical resources.

SR-IOV-capable devices can advertise several PFs each of which can support multiple VFs.
The Virtual Machine Monitor (VMM) allows assigning one or more VFs to a VM, mapping the VF configuration space into the VM.
Memory translation technologies usually relying on the I/O Memory Management Unit (IOMMU) such as Intel VT-d \cite{intelvtd} or AMD-Vi \cite{amdvi}, provide a hardware-based solution to enable virtualization of I/O resources including DMA and interrupt remapping, allowing direct transfers between the device and the VM.

\subsection{KVM, QEMU, libvirt, and VFIO}

Kernel-based Virtual Machine (KVM) is an open-source Linux subsystem that implements virtualization capabilities in a form of a VMM (or hypervisor) \cite{kvm}.
It provides a complete virtualization solution, including virtualized CPU, memory, devices, and I/O, delivering near-native performance thanks to the use of hardware virtualization extensions.
KVM allows multiple virtual machines to run on a single physical host, using a modular design with a frontend driver for managing guests' I/Os and a backend driver for managing and multiplexing VMs' I/O to the real device.

QEMU is an open-source emulator that provides a set of devices that can be used in the guest systems, also implementing the backend drivers for KVM.
It is used in conjunction with KVM with the help of libvirt, a toolkit that provides a common API for managing VMs and their resources to simplify their integration and management.

Libvirt, a library that provides high-level API for managing virtualization technologies, including QEMU, is used to simplify and abstract the interface to manage VMs, providing additional functionalities for resource access and management without sacrificing performance.
The library can interact with QEMU through the QEMU Monitor Protocol (QMP) for controlling the VM, through the QEMU Guest Agent for communicating with the guest, and optionally via a remote display protocol for connecting to a VM graphical console.
QMP enables control of the VM state, configuration of virtual devices eventually hotplugging them, performance monitoring, and snapshot management.
For device control, it offers functionalities for adding, removing, and configuring PCI devices.

QEMU also implements Virtual Function I/O (VFIO), a framework that takes the place of the device driver in the KVM host, reserving host resources and providing access to them on behalf of the guest.
It provides pass-through functionalities such as direct access to the device using the IOMMU, for the highest possible I/O performance and security through isolation of devices of the same IOMMU group.
VFIO is often used in cloud computing, high-performance computing, and edge computing, where it is necessary to provide VMs with direct access to hardware devices in order to have high performance and low latencies.

\section{Related Work}

Academic studies on the virtualization of FPGA resources are abundant in the research community \cite{futurefpga, fpga_analysis}, covering a wide range of FPGA types (including PCIe, network, and SoC) and methodologies.

The paper \cite{enabling} proposes a framework where a control node is used to handle requests from tenants, schedule resources, and allocate and control VMs.
The pvFPGA \cite{pvfpga} is a hardware solution that uses the Xen hypervisor and a paravirtualized environment to enable multiple domains to access the FPGA resources.
This requires the backend driver on domain 0 to manage all the traffic between the VMs and the custom DMA inside the FPGA.
FPGAVirt \cite{fpgavirt} is a hardware-software solution that enables the communication between guest and FPGA using virtio to create a paravirtualized framework and an FPGA overlay allowing FPGA re-configuration and re-allocation of resources (VF).
In particular, the HW overlay used to give flexibility is also in charge of the isolation and sandboxing, a crucial part that is left to SR-IOV in the SVFF.
vFPGAmanager \cite{vfpgamanager} is a hardware-software framework for exploiting FPGA resources in a Network Function Virtualization (NFV) environment.

The SVFF framework takes into account the need for modifying the VF configuration, and is generic enough to handle various types of VFs, whereas most current solutions for SR-IOV either focus solely on network virtualization \cite{virtnet,pciedma,highperfnet,raccoon,srvm}, storage \cite{fvm} or lack these capabilities.

The framework proposed in this paper:
\begin{enumerate*}[label=(\arabic*)]
    \item is an application-generic solution that requires no driver modification on the guest;
    \item uses the well-tested SR-IOV implemented in the QDMA, requiring the user to just attach a digital device (FPGA implemented) to the DMA IP;
    \item automatize all the FPGA, device, driver, and virtualization management operations;
    \item provides the pause functionality.
\end{enumerate*}

\section{SVFF: Design and Implementation}

This chapter describes the design and implementation of the SVFF framework, illustrating in detail the specific components and features it implements.
The framework is a virtualization solution that provides a way to manage VFs on PCIe-attached FPGA devices.
It acts as an intermediary between the VMM and the device driver and provides commands for automatically initializing and reconfiguring the FPGA device.

Section \ref{qdma_ip} will focus on the hardware component, based on Xilinx QDMA, which provides SR-IOV support on a PCIe-attached FPGA, and serves as the hardware virtualization support, in gray in figure \ref{fig:sriov}.

Section \ref{software} will cover the core of the SVFF virtualization solution, the part centered around the software component, which leverages VFIO on QEMU/KVM and libvirt for automatic VF management to implement the \textit{init} and \textit{reconf} functionalities, in blue in figure \ref{fig:sriov}.
It will also address the integration with QEMU/KVM and libvirt environment, in red in figure \ref{fig:sriov}.

\subsection{SVFF: the hardware requirement}\label{qdma_ip}

In general, the key hardware requirement considered while developing SVFF was the use of a standard solution to enable HW programmers to provide accelerators that can be easily virtualized using SR-IOV.
As a result, the SVFF hardware requirements are simple: the Xilinx Queue-based Direct Memory Access (QDMA) Intellectual Property (IP) needs to be integrated into the system bitstream.

More in detail, QDMA is an IP core that implements a DMA engine optimized for high bandwidth and high packet count data transfers, internally composed of the UltraScale+ Integrated Block for PCI Express and an extensive DMA and bridge infrastructure.
It supports PCIe devices up to Gen3x16, up to a 2048 queue set, polling or interrupt mode, and provides the necessary SR-IOV functionality.
The QDMA IP supports SR-IOV and the creation of up to 4 PFs and 252 VFs, providing a flexible configuration mechanism for the VFs through the Linux drivers.
On the FPGA side, the QDMA supports the UltraScale+ devices and provides both AXI4 Memory Mapped and AXI4 Stream interfaces for connecting to other FPGA-implemented devices.

\subsection{SVFF: the software framework}\label{software}

The framework automates the initialization, creation, and configuration of VFs, leveraging the Xilinx QDMA Linux kernel driver.
More in particular, the SVFF is composed of several modules that have been developed to assist in the management of the QDMA host driver (the QDMA manager) and perform flash operations using Xilinx Vivado or the Xilinx Software Command-Line Tool (XSCT).
Hereafter, the main components of the framework are described, starting with the novel pause functionality, followed by a description of the QEMU/KVM libvirt integration, and software automation.

\subsubsection{Pause Functionality} \label{pause}

The main drawback of SR-IOV is that it requires the removal of all the VFs by setting the number of VF for a PF to zero before changing it.
To overcome this limitation, the pause functionality has been developed to provide a way to temporarily detach a VFIO device connected to a VM from the host side only.
In this way, the removal of the device from the guest is avoided without causing interruption of the services due to driver unloading.
This solution acts only on the host system, leaving complete flexibility on the guest that can use un-modified drivers.

Since the VFIO device class implemented in QEMU only provides functionalities for creating (\texttt{realize}) and destroying (\texttt{exit}) a VFIO device type, the class has been extended with the pause function.
The pause functionality allows to pause or unpause the device based on its status and its pausability (it is active and tested only for Xilinx devices but should work on every PCI device, eventually with minor tweaks).

\begin{figure}
\centering
\includegraphics[width=\columnwidth]{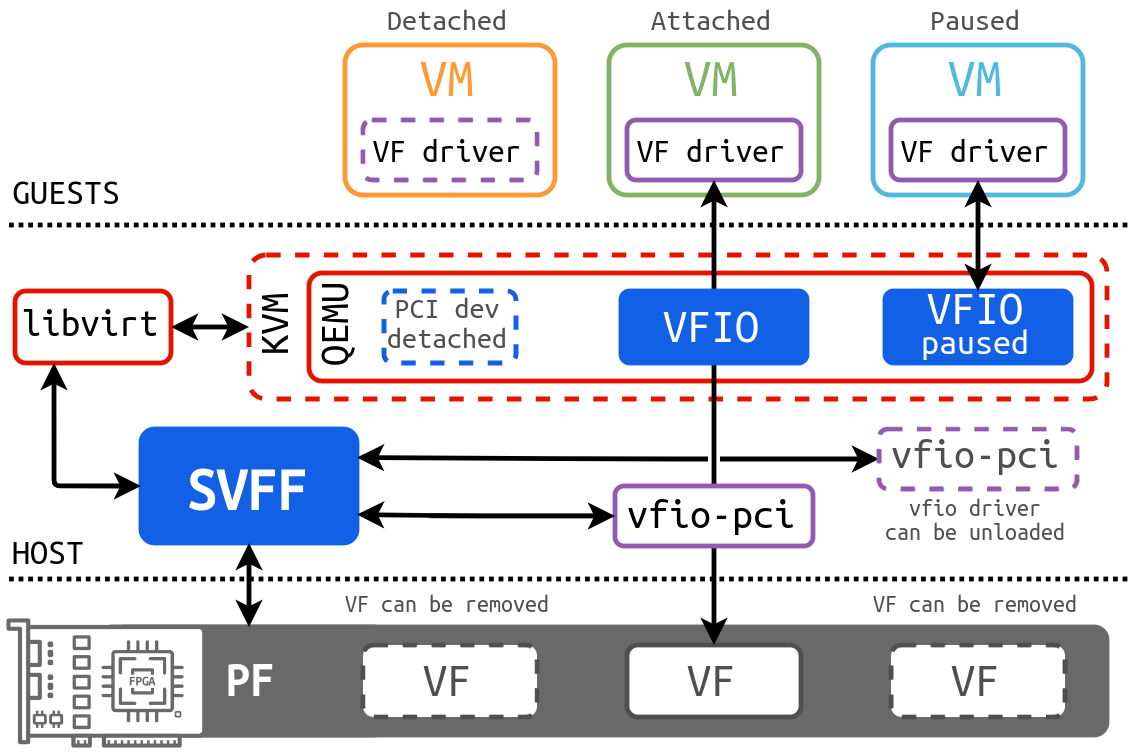}
\caption{Device detached (left), attached (center), and paused (right)}
\label{fig:pause}
\end{figure}

The \textbf{pause} procedure of a VFIO device consists of 3 main steps:
\begin{enumerate}
    \item save the PCI device config space including emulated config space and Message Signaled Interrupts (MSI) state to be restored in the unpause;
    \item perform unregister operations on the PCI device, including deletion of PCI memory subregions, optional device ROM, and interrupt bits;
    \item perform unregister operations on the VFIO device, including un-registration of device request and error notifier (requests from the guest to the device will be ignored), disabling and teardown of interrupts, deletion of VFIO device BARs and exiting from IOMMU group (this is needed to completely detach the device from the host).
\end{enumerate}
This procedure will effectively detach the device from the host, keeping it visible but in a paused state in the VM, as can be seen in figure \ref{fig:pause}.
In this state, the guest will still be able to see the device and its emulated registers but can not do any actual I/O operations until the device is unpaused.

The \textbf{unpause} procedure of a previously paused VFIO device consists of 2 main steps:
\begin{enumerate}
    \item restore the I/O connections of the VFIO device registering again the BARs, adding the device to the IOMMU group, and registering the interrupt notifiers without affecting the emulated registers;
    \item restore the PCI device config registers to the values saved before the pause, eventually updating the memory regions mappings.
\end{enumerate}
This procedure will re-attach the VFIO device to the guest, enabling again I/O operations.

\subsubsection{Integration with QEMU and libvirt}

The framework integrates with QEMU and libvirt to provide a streamlined and flexible mechanism for managing VFs in a virtualized environment.
In order to provide controls for the additional pause functionality, QEMU has been modified to add a new \texttt{paused} device's property that provides callbacks on its modification.
A new command has been registered in the QEMU Monitor Protocol (QMP) to allow sending to QEMU a \texttt{device\_pause} command with as first argument the device id and as second the pause status.
When the command is executed, if the device class provides a \texttt{pause()} function, the monitor will call it to execute the action provided by the device.
The QEMU vfio-pci implementation (under \texttt{hw/vfio/pci.c}) has been modified to introduce the pause status and the callbacks to pause and unpause the device as described in section \ref{pause}.

When the pause functionality is enabled, the attach and detach operations will behave differently.
During the detachment, if the VM is live, the framework will send the QMP pause command for the device instead of asking to detach. It will then unbind the VF from the VFIO driver.
For the attachment, if the VM is live and the VF was previously paused, the framework will bind the VF to the VFIO driver using the QDMA Manager and then send the QMP unpause command.

\subsubsection{Automations}

Software automation is crucial in server environments for HPC, cloud computing, edge computing, and other use cases that involve repetitive tasks and minimal user interactions. In order to provide a comprehensive solution for users, SVFF handles various tasks including FPGA bitstream flashing, QDMA driver interactions, and VF attachment and detachment.

The SVFF provides two main functionalities for the user: the \textit{init} function used to initialize the FPGA device the first time is connected to the host, eventually flashing the bitstream, and the \textit{reconf} function used to change the number of VFs to be created and attached to the VMs.

The \textit{init} process involves a recursive search for all the VFs associated with the PFs of the device, detaching them from the VM, and removing the PF from the PCI bus unloading its driver.
Afterward, the bitstream can be flashed to the FPGA (automation of the bitstream flash procedure is achieved using TCL scripts for Xilinx Vivado or XSCT), the PF is discovered and configured, and the number of queues and VFs to be used is set.
The new VFs are then attached to the VMs, making them available for the guests, and will use the \texttt{vfio-pci} as the backend driver on the host and the \texttt{qdma-vf} driver on the VM, leaving to QEMU the management of the passthrough operations.
On the host system, the \texttt{qdma-pf} driver provides a sysfs interface to configure and control various parameters of the PF exposed by the QDMA IP.

After the initial number of VFs is set using the init option, there may be a need to change the configuration using the \textit{reconf} command.
This could involve removing a VF from a VM, adding more VFs to a new or existing VM, or reflashing the FPGA bitstream to change the HW configuration.
The reconfiguration option is used to change the number of VFs for a given PF and automatically manage the detach, reattach, and pause of a VF from its related VM.

Attaching and detaching Virtual Function devices to VMs is automated in the program and facilitated through the use of virsh, which serves as the primary interface for managing guest domains with libvirt.
In the attach process, each VF device is specified in an XML file that outlines its properties, such as device type, address, and driver.
This file is saved to maintain a record of the VF-VM association for future reference, allowing for a seamless detach operation.
Taking into account whether the domain is live (running) or not, virsh is then used to handle the actual attachment of the VF to the VM.
For the detachment process, the properties of the VF and the VM it is attached to are retrieved from the corresponding XML file, and virsh is utilized to detach the device from the VM.

Finally, interactions with the QDMA driver are mediated with the QDMA manager, a program developed to simplify the management of the drivers offering various functionalities, such as unbinding a PCI device from its driver, binding the VFIO driver to it, removing the device and its related VFs from the PCI bus, and rescanning the PCI bus to detect new devices (which is useful when initializing the FPGA after it has been attached to the host).
It also includes security checks for the device ID and driver name, a recursive search for related VFs, and management of new device IDs for VFIO.

\section{Evaluation and Results} \label{evaluation_and_results}

In this section, the evaluation methodology and the results of the experiments conducted on the proposed framework are presented.
Since the SVFF leverages the same underlying SR-IOV technology provided by the QDMA introducing significant changes only at a software level, I/O performance measurements are left out as they would reflect the Xilinx QDMA reference values \cite{qdmaperf}.

Consequently, the purpose of the evaluation is to assess the performance and overhead of the SVFF novel pause functionality.
In this context, there are two scenarios: the standard SR-IOV detach-attach method on one side, and the newly introduced SVFF pause-unpause.
The time needed to complete an SR-IOV re-configuration cycle is measured in both cases, aiming at assessing the overhead introduced by SVFF.
More in detail, the measured re-configuration cycle assumes that a number of VFs is already attached to the same number of VMs (one VF per VM) and measures the time it takes to remove/pause all the VFs and attach/unpause them again to the corresponding VMs.

Delays are measured from the host side since guests are unaware of when a VF is paused/unpaused using the default QDMA guest driver.

\subsection{FPGA Design and Bitstream}

As mentioned above, nothing more than the QDMA IP is needed to enable the framework, leaving extreme flexibility to the user to implement custom blocks in the FPGA for any kind of application.

For this paper's benchmarks, the Xilinx Vivado 2022.1 toolset was used to generate the bitstream for a Xilinx Alveo U55C target.
The FPGA design is a Vivado block design with the QDMA IP v4.0 connected to two BRAM controllers, one via the AXI and one via the AXI Lite interface, each with its corresponding memory.
Additionally, a custom Function Level Reset (FLR) block implemented in VHDL has been included to assert FLR requests without actually performing any reset action.
This is required by the QDMA which would otherwise wait indefinitely for the FLR response to arrive.
The QDMA IP is configured with an 8 GT/s x16 PCIe link with a 512-bit AXI data interface and 512 queues.
SR-IOV is also enabled providing 1 PF classified as a memory controller with 32 VF.

From the host's perspective, the PCIe PF and each VF that inherits its properties expose two memories: a fast 512KB used to mimic memory accesses and a slow 32KB for device-like operations.

\subsection{Software Configuration}

The host system is a 32 cores, 128GB RAM x86\_64 server running Red Hat Enterprise Linux (RHEL) 8.7 with Linux kernel 4.18 and hosting the Alveo FPGA card, managed with the Xilinx Runtime library (XRT) 2.13.466 branch 2022.1.
Qemu 7.1.0 has been modified to add the pause functionality, compiled with GCC 8.5.0 and the \texttt{--x86\_64-softmmu} and \texttt{--enable-kvm} parameters, and installed in the \texttt{/usr/local} prefix.
SELinux is used in enforcing mode on the host, requiring the installation of QEMU binaries to the \texttt{qemu\_exec\_t} context.
Libvirt 8.0.0 is used and the kernel is started with the \texttt{intel\_iommu=on} command line option to enable the IOMMU.

Each guest VM is a 2 cores, 2GB RAM x86\_64 Ubuntu 22.04 with Linux kernel 5.15 based on a Q35+ICH9 machine.

The QDMA drivers are compiled from the source and installed in both the host and the guest systems, which will use only the PF and the VF parts respectively.

\subsection{Results}

\begin{table}
\renewcommand{\arraystretch}{1.3}
\caption{VF detach-attach vs pause-unpause overhead, AVG of 100 runs}
\label{tab:overhead}
\centering
\begin{tabular}{|l|rr|rr|cc|}
\hline
\multirow{2}{*}{\textbf{\#VF}} & \multicolumn{2}{c|}{\textbf{Detach/Attach}}          & \multicolumn{2}{c|}{\textbf{Pause/Unpause}}          & \multicolumn{2}{c|}{\textbf{Overhead}} \\
                               & \multicolumn{1}{c}{avg ms} & \multicolumn{1}{c|}{$\sigma$} & \multicolumn{1}{c}{avg ms} & \multicolumn{1}{c|}{$\sigma$} & \%                 & ms/VF             \\ \hline
1                              & 4151                       & 40                      & 4068                       & 56                      & -2.00              & -83               \\ \hline
4                              & 12988                      & 183                     & 12665                      & 171                     & -2.49              & -80               \\ \hline
10                             & 31129                      & 497                     & 30285                      & 505                     & -2.71              & -84               \\ \hline
\end{tabular}
\end{table}

Table \ref{tab:overhead} shows the average delays on 100 runs for both configurations when operating on 1, 4, and 10 VF on as many VMs.
The overhead values show that there is a minimal reduction in the delays from 2\% to 2.71\% with a reducing time of around 80ms per VF in all the scenarios.
The delays are not linearly increasing with the number of VF, meaning that some operations are not dependent on it.

To understand better the results, the duration of the different macro-operations performed in the two cases for a single run with 1, 4, and 10 VFs are inspected,  showing the values in table \ref{tab:compare}.
It's important to note that these timings represent one particular run, and the average values may differ across multiple runs or in different environments.
The rescan operation of the PCI bus is performed to be sure to have discovered all the PF and VF on the bus, and its time is the same in both the Detach/Attach (D/A) and Pause/Unpause (P/U), not dependent on the number of VF involved.
The removal (detach or pause) of the VF in step 2 takes on average the same time in both scenarios, although the pause skips some device removal operations performed by detach but spends some more time allocating and saving the device configuration data structures.
Step 3 shows the delay for changing the number of VFs of a PF and it is quite similar in both cases, slightly increasing with the number of VF involved.
In step 4, it is noticeable the main gain of the unpause operation with respect to the attach. This is justified by the fact that the former skips some of the realize operations of the device (because it does not detach it from the guest side), copying back the device configuration data instead.

\begin{table}
\renewcommand{\arraystretch}{1.3}
\caption{Timings of VF detach-attach and pause-unpause operations}
\label{tab:compare}
\centering
\begin{tabular}{|l|rr|rr|rr|}
\hline
\multirow{3}{*}{\textbf{Step}} & \multicolumn{2}{c|}{\textbf{1 VF}}                                   & \multicolumn{2}{c|}{\textbf{4 VFs}}                                  & \multicolumn{2}{c|}{\textbf{10 VFs}}                                 \\
                               & \multicolumn{1}{c}{\textbf{D/A}} & \multicolumn{1}{c|}{\textbf{P/U}} & \multicolumn{1}{c}{\textbf{D/A}} & \multicolumn{1}{c|}{\textbf{P/U}} & \multicolumn{1}{c}{\textbf{D/A}} & \multicolumn{1}{c|}{\textbf{P/U}} \\
                               & \multicolumn{1}{c}{ms}           & \multicolumn{1}{c|}{ms}           & \multicolumn{1}{c}{ms}           & \multicolumn{1}{c|}{ms}           & \multicolumn{1}{c}{ms}           & \multicolumn{1}{c|}{ms}           \\ \hline
1. rescan                      & 138                              & 138                               & 144                              & 141                               & 139                              & 139                               \\ \hline
2. remove VF                   & 1265                             & 1273                              & 5417                             & 5505                              & 14360                            & 13878                             \\ \hline
3. change \#VF                 & 1256                             & 1295                              & 1460                             & 1412                              & 1817                             & 1730                              \\ \hline
4. add VF                      & 1472                             & 1346                              & 5946                             & 5653                              & 15042                            & 14448                             \\ \hline
\textbf{total}                 & \textbf{4131}                    & \textbf{4052}                     & \textbf{12967}                   & \textbf{12711}                    & \textbf{31358}                   & \textbf{30195}                    \\ \hline
\end{tabular}
\end{table}

It is also to mention that the timings reported are real timings (the actual time elapsed from the start to the end of the operation) and are mostly dependent on the SR-IOV operations performed on the hardware: for a 4 s reconfiguration of 1 VF, CPU time (user + kernel) is only around 350 ms, less than 9\% (this stands also for the 4 VFs and 10 VFs cases tested).

This benchmark analysis demonstrated that the pause functionality can be used as a suitable alternative to device removal, as it incurs no significant overhead and can even provide a slight performance improvement of a few percentage points, with the main advantage of avoiding device removals on the guest side.

\section{Conclusion and future work}

This paper presented the SR-IOV Virtual Function Framework (SVFF), a virtualization solution that provides a way to manage Virtual Functions on PCIe-attached FPGAs devices.
The framework simplifies the management of VF by automating the assignment of VFs to different VMs also allowing re-configuration on the fly.
The pause functionality allows for seamless VF reconfiguration from the VM side, where the guest does not see the device removed but is only unavailable for the process duration.
Moreover, this approach slightly reduces the delay time by over 2\%.
Overall, the SVFF solution provides an efficient and effective method for managing FPGA resources in virtualized environments, particularly in data centers, cloud computing, and HPC environments, where device management can be a challenging task.

Future work will focus on optimizing the performance of SVFF and implementing higher-level features.
One area for improvement is the pause functionality, which could be optimized to enable faster recovery times and minimal interruption to FPGA-based workloads, also keeping track of the guest driver requests that are currently ignored when the device is paused.
Another potential area for future work is the implementation of dynamic resource allocation for FPGAs based on workload demands to improve the efficiency of FPGA utilization and enable users to allocate and deallocate FPGA resources in real-time,
enhancing the overall performance of cloud-based applications.

\section*{Acknowledgment}
This work has received funding from the EU Horizon 2020 Programme, grant agreement No 957269 (EVEREST) \cite{everest}.

\bibliographystyle{IEEEtran}
\bibliography{svff}

\end{document}